\documentclass[traditabstract]{aa}

\usepackage{graphicx}
\usepackage{txfonts}
\usepackage{natbib}
\bibpunct{(}{)}{;}{a}{}{,}

\def\nul#1{}
\def\be \begin{equation}
\def\ee \end{equation}
\def\figpath{}

\begin{document}

\title{The HARPS search for southern extra-solar planets\thanks{Based
    on observations made with the HARPS instrument on the ESO 3.6 m
    telescope at La Silla Observatory under the GTO programme ID
    072.C-0488. The table with the radial velocities is available in 
    electronic form at the CDS via anonymous ftp to cdsarc.u-strasbg.fr
   (130.79.128.5) or via http://cdsweb.u-strasbg.fr/cgi-bin/qcat?J/A+A/????  
}  }
   \subtitle{XVI. {HD}\,45364, a pair of planets in a 3:2 mean motion resonance}
   \titlerunning{{HD}\,45364, a pair of planets in a 3:2 mean motion resonance}

\author{
A.C.M. Correia\inst{1}
\and S.~Udry\inst{2}
\and M.~Mayor\inst{2}
\and W.~Benz\inst{3}
\and J.-L.~Bertaux\inst{4}
\and F.~Bouchy\inst{5}
\and J.~Laskar\inst{6} 
\and \\ C.~Lovis\inst{2}
\and C.~Mordasini\inst{3}
\and F.~Pepe\inst{2}
\and D.~Queloz\inst{2}
}

\offprints{A.C.M. Correia}
 
\institute{Departamento de F\'isica, Universidade de Aveiro, Campus de
  Santiago, 3810-193 Aveiro, Portugal \\
  \email{correia@ua.pt}
  \and Observatoire de Gen\`eve, Universit\'e de Gen\`eve, 51 ch.
  des Maillettes, 1290 Sauverny,  Switzerland \\
  \email{michel.mayor@obs.unige.ch}
  \and Physikalisches Institut, Universit\"at Bern, Silderstrasse 5, 
CH-3012 Bern, Switzerland
  \and Service d'A\'eronomie du CNRS/IPSL, Universit\'e de Versailles
  Saint-Quentin, BP3, 91371 Verri\`eres-le-Buisson, France 
  \and Institut d'Astrophysique de Paris, CNRS, Universit\'e Pierre et
  Marie Curie, 98bis Bd Arago, 75014 Paris, France 
  \and IMCCE, CNRS-UMR8028, Observatoire de Paris, UPMC, 77 avenue
  Denfert-Rochereau, 75014 Paris, France 
}

\date{Received ; accepted To be inserted later}

  \abstract{Precise radial-velocity measurements with the {\small HARPS} spectrograph
  reveal the presence of two planets orbiting the solar-type star {HD}\,45364.
  The companion masses are $m \sin i = 0.187\,M_{\rm Jup}$ and $0.658\,M_{\rm Jup}$,
  with semi-major axes of $a=0.681$\,AU and $0.897$\,AU, and eccentricities of
  $e=0.168$ and $0.097$, respectively. A dynamical analysis of the system further
  shows a 3:2 mean motion resonance between the two planets, which prevents close
  encounters and ensures the stability of the system over 5\,Gyr. This is the first
  time that such a resonant configuration has been observed for extra-solar planets,
  although there is an analogue in our Solar System formed by Neptune and Pluto. This
  singular planetary system may provide important constraints on planetary formation
  and migration scenarios.}


   \maketitle
%


\section{Introduction}

At present, about 25\% of the known exoplanets are in 
multi-planetary systems. The older the radial-velocity planet-search 
surveys, the higher the fraction of multi-planet families detected in these programs. 
Taking into account the still strong bias against long-period and/or low-mass planet 
detection, and the enhanced difficulty of fully characterizing systems with more than one
planet, the fraction of known multi-planet systems is certainly still a lower limit.
It appears likely that a high number, if not the majority, of planet-host stars form 
systems of planets rather than isolated, single planetary companions.

Among the known multi-planet systems, a significant fraction are in mean motion
resonances, the majority of which are in low-order resonances. The 2:1 resonance is
the most common (HD\,73526, HD\,82943, HD\,128311, GJ\,876), but other configurations 
are observed as well, such as the 3:1 resonance in HD\,75732 or the 5:1 resonance in 
HD\,202206. These resonances most probably arise from evolutionary processes as 
migrating planets forming in the protoplanetary disc become trapped. 
In our Solar System, we also find mean motion resonances in the satellites of
the giant planets, or between these planets and several asteroids.
The most well-known example is the Io-Europa-Ganymede system in a 4:2:1
resonance, or the Neptune-Pluto system in a 3:2 resonance.
While the satellites are believed to achieve resonant configurations after tidal
evolution of their orbits, the resonances between planets and asteroids were
probably formed after the inward or outward migration of the planets during
the early stages of the evolution of the Solar System.

The presence of two or more interacting planets in a system dramatically increases 
our potential ability to constrain and understand the processes of planetary formation 
and evolution.
The dynamical analysis of such systems is then very useful, first for
constraining the system evolution history and second for determining the system
``structure'' in terms of orbital content.  

Multi-planet systems are naturally found in planet-search programs. However, improving the
precision of the radial-velocity measurements greatly helps their detection and complete 
characterization. The {\it {\small HARPS} search for southern extra-solar planet} is an extensive 
radial-velocity survey of some 2000 stars of the solar neighborhood conducted with the HARPS 
spectrograph on the ESO 3.6-m telescope at La\,Silla (Chile) in the framework of the Guaranteed 
Time Observations granted to the HARPS building consortium \citep{Mayor_etal_2003}. About half 
of the HARPS GTO time is dedicated to very high-precision measurements of non-active 
stars selected from the CORALIE planet-search program  \citep{Udry_etal_2000} and stars with 
already known giant planets, while searching for lower mass companions. This program 
reveals itself as very efficient in finding multi-Neptune \citep{Lovis_etal_2006} and multi-Super Earth 
systems \citep{Mayor_etal_2008}.

From the HARPS high-precision survey, we present an interesting system of 
two planets in a 3:2 mean motion resonance around HD\,45364, a configuration not observed 
previously among extra-solar planet, but similar to the Neptune-Pluto pair in the
Solar System. 
Section\,2 gathers useful stellar information about HD\,45364, its derived orbital solution is described 
in Sect.\,3, and the dynamical analysis of the system is discussed in Sect.\,4. Finally, conclusions 
are drawn in Sect.\,5.


\section{Stellar characteristics of HD\,45364}

The basic photometric (K0V, $V$\,=\,8.08, $B$\,$-$\,$V$\,=\,0.72)
and astrometric ($\pi$\,=\,30.69\,mas) properties of HD\,45364 were
taken from the Hipparcos catalogue \citep{Esa_1997}. They are recalled
in Table\,\ref{TableStar} with inferred quantities such as the
absolute magnitude ($M_V$\,=\,5.51) and the stellar physical
characteristics derived from the HARPS spectra by \citet{Sousa_etal_2008}.
For the complete high-precision HARPS sample (including HD\,45364), these authors
provided homogeneous estimates of effective temperature  
($T_{\rm eff}$\,=\,5434\,$\pm$\,20\,K), metallicity ([Fe/H]\,=\,$-0.17$\,$\pm$\,0.01), 
and surface gravity ($\log{g}$\,=\,4.38\,$\pm$\,0.03) of the stars. A projected
low rotational velocity of the star, $v\sin{i}=1$\,kms$^{-1}$ was derived from a 
calibration of the width of the cross-correlation function used in the 
radial-velocity estimate \citep{Santos_etal_2002}.

HD\,45364 is a non-active star in our sample with an activity indicator 
$\log R^{\prime}_{HK}$ of $-4.94$. No significant radial-velocity jitter is thus 
expected for the star.  From the activity indicator, we also derive a stellar 
rotation period $P_{rot}$\,=\,32\,day \cite[following ][]{Noyes_etal_1984}.

HD\,45364 has a subsolar metallicity with [Fe/H]\,=\,$-0.17$ unlike most of the
gaseous giant-planet host stars \citep{Santos_etal_2004}.  According to simulations 
of planet formation based on the core-accretion paradigm, moderate 
metal-deficiency does not, however, prevent planet formation 
\citep{Ida_Lin_2004,Mordasini_etal_2008}. Taking into account its subsolar metallicity, 
\citet{Sousa_etal_2008} derived a mass of 0.82\,M$_{\odot}$ for the star.  
From the color index, the derived effective temperature, and the corresponding 
bolometric correction, we estimated the star luminosity to be 0.57\,L$_\odot$.

\begin{table}
\caption{Observed and inferred stellar parameters of HD\,45364. 
\label{TableStar} }
\begin{center}
\begin{tabular}{l l c c}
\hline\hline
\multicolumn{2}{l}{\bf Parameter} &\hspace*{2mm} & \bf HD\,40307 \\
\hline
Sp & & & K0\,V  \\
$V$ & [mag] & & 8.08 \\
$B-V$ & [mag] & & 0.72  \\
$\pi$ & [mas] & & 30.69 $\pm$ 0.81 \\
$M_V$ & [mag] & & 5.51 \\
$T_{\mathrm{eff}}$ & [K] & & 5434 $\pm$ 20 \\
log $g$ & [cgs] & & 4.38 $\pm$ 0.03 \\
$\mathrm{[Fe/H]}$ & [dex] & & $-0.17$ $\pm$ 0.01 \\
$L$ & [$L_{\odot}$] & & 0.57 \\
$M_*$ & [$M_{\odot}$] & & 0.82 $\pm$ 0.05 \\
$v\sin{i}$ & [km s$^{-1}$] & & $1$ \\
$\log R'_{\mathrm{HK}}$ & & & $-4.94$ \\
$P_{\mathrm{rot}}$($\log R'_{\mathrm{HK}}$) & [day] & & 32 \\
\hline
\end{tabular}
\end{center}
Photometric and astrometric data are from the Hipparcos catalogue
  \citep{Esa_1997} and the stellar physical quantities from
  \citet{Sousa_etal_2008}.\end{table}


\section{Orbital solution for the HD\,45364 system}
\label{orbsolutions}

\begin{table}
\caption{Orbital parameters for the two bodies orbiting {HD}\,45364, obtained
with a 3-body Newtonian fit to observational data (Fig.\,\ref{F1}).  \label{T2} }
\begin{center}
\begin{tabular}{l l c c} \hline \hline
\noalign{\smallskip}
{\bf Param.}  & {\bf [unit]} & {\bf {HD}\,45364\,b} & {\bf {HD}\,45364\,c} \\ \hline 
\noalign{\smallskip}
Date         & [JD-2400000]         & \multicolumn{2}{c}{53500.00 (fixed)}  \\ 
$V$          & [km/s]               & \multicolumn{2}{c}{$ 16.4665 \pm 0.0002 $}  \\  
$P$          & [day]                & $ 226.93 \pm 0.37  $ & $ 342.85 \pm 0.28  $ \\ 
$\lambda$    & [deg]                & $ 105.76 \pm 1.41  $ & $ 269.52 \pm 0.58  $ \\ 
$e$          &                      & $ 0.1684 \pm 0.0190 $ & $ 0.0974 \pm 0.012 $ \\ 
$\omega$     & [deg]                & $ 162.58 \pm 6.34  $ & $ 7.41 \pm 4.30  $ \\ 
$K$          & [m/s]                & $ 7.22 \pm 0.14  $ & $ 21.92 \pm 0.43  $ \\  
$i$          & [deg]                & $ 90 $ (fixed)      & $ 90 $ (fixed)      \\  \hline
\noalign{\smallskip}
$a_1 \sin i$ & [$10^{-3}$ AU]       & $ 0.1485 $           & $ 0.6874 $ \\
$f (M)$      & [$10^{-9}$ M$_\odot$]& $ 0.0085  $           & $ 0.3687  $ \\
$M \sin i$ & [M$_\mathsf{Jup}$]   & $ 0.1872 $           & $ 0.6579 $ \\
$a$          & [AU]                 & $ 0.6813 $           & $ 0.8972 $ \\ \hline
\noalign{\smallskip}
$N_\mathrm{meas}$        &          & \multicolumn{2}{c}{58}   \\
Span         & [day]                & \multicolumn{2}{c}{1583}   \\
$rms$        & [m/s]                & \multicolumn{2}{c}{1.417}   \\
$\sqrt{\chi^2}$     &                      & \multicolumn{2}{c}{2.789}   \\ \hline
\end{tabular}
\end{center}
Errors are given by the standard deviation $ \sigma $ and $ \lambda $ is the
mean longitude of the date ($ \lambda = \omega + M $). 
The orbits are assumed co-planar. 
\end{table}

\begin{figure}
   \centering
    \includegraphics*[height=8.5cm,angle=270]{\figpath 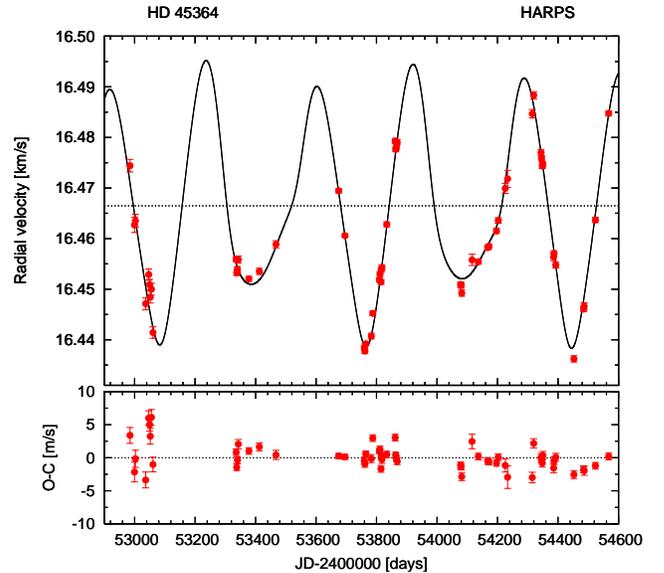}
  \caption{{\small HARPS} radial velocities for {HD}\,45364, 
    superimposed on a 3-body Newtonian orbital solution (Table\,\ref{T2}).
    \label{F1} }
\end{figure}

The {\small HARPS} observations of {HD}\,45364 started in December 2003 
and have now been going for about four years and a half. From the first stages of 
the observations, peculiar variations in the radial velocities (Fig.\,\ref{F1})
have shown the presence of one or more companions in the system. 
After 58 measurements, we are now able to determine the nature of these bodies.
Using a genetic algorithm combined with the iterative Levenberg-Marquardt method
\citep{Press_etal_1992}, we first attempted to fit the complete set of radial velocities
using a model with two Keplerian orbits.
This fit yields an adjustment of $\sqrt{\chi^2}=2.71$ and $rms=1.38\,\mathrm{m
s}^{-1}$ with two planets, one at $P=225.8$\,day, $e=0.174$, and a minimum mass of
$0.186\,M_\mathrm{Jup}$, and another at $P=343.9$\,day, $e=0.097$, and a minimum 
mass of $0.659\,M_\mathrm{Jup}$. 

\begin{figure}
   \centering
    \includegraphics*[height=8.5cm,angle=270]{\figpath 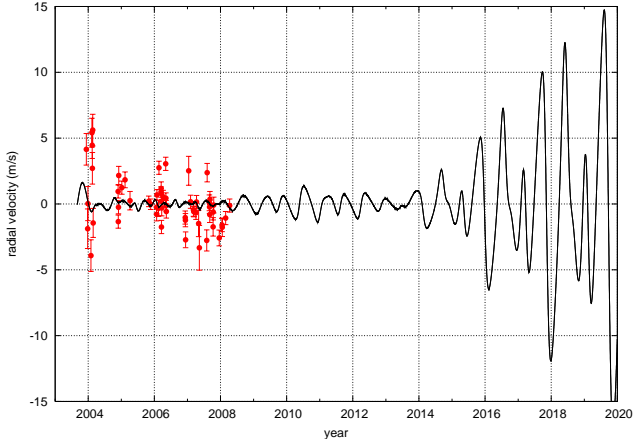}
  \caption{Radial velocity differences between the two independent 
    Keplerian model and the 3-body Newtonian model
    (Table\,\ref{T2}).  Data coincides at JD=2453776 (Feb. 9$^\mathrm{th}$ 2006). 
    We also plot the velocity residuals of the 3-body fit (Fig.\,\ref{F1}).
    With the current {\small HARPS}'s precision for this star, we expect
    to observe these differences within ten years.  \label{F2}}
\end{figure}

Due to the proximity of the two planets and their high minimum masses, the
gravitational interactions between these two bodies are strong.  
This prompts us to fit the observational data using a 3-body Newtonian model,
assuming co-planar motion perpendicular to the plane of the sky, similarly to
what has been done for the system {HD}\,202206 \citep{Correia_etal_2005}. 
The orbital parameters corresponding to the the best fitted solution are listed
in Table\,\ref{T2}. 
We get identical results for $ \sqrt{\chi^2} $ and velocity
residuals as obtained with the two-Keplerian fit. 
Although there is no significant improvement in the fit, an important difference
exists: the new orbital parameters for both planets show some 
deviations from the two-Keplerian case. In particular, the arguments of the
periastrons ($ \omega $) show a difference of some degrees.
We then conclude that, despite being unable to detect the
planet-planet interaction in the present data, the orbits undergo important
perturbations, and we expect to detect this gravitational
interaction in the future. In Fig.\,\ref{F2}, we plot the two best fit models
evolving in time and clearly observe detectable deviations between the
two curves that appear within ten years.
The 3-body Newtonian fit provides a more accurate approximation of the
{HD}\,45364 planetary system, and the orbital parameters thus determined will be
adopted as reference henceforward (Table\,\ref{T2}). 

We also fitted the data with a 3-body Newtonian model for which
the inclination of the orbital planes, as well as
the node of the outer planet orbit, were free to vary. 
We were able to find a wide variety of configurations, some with low inclination values
for one or both planets, that slightly improved our fit to a minimum $
\sqrt{\chi^2} = 2.32 $ and $rms$\,=\,$1.14\,\mathrm{m/s}$. 
However, all of these determinations are uncertain, and since we also increase
the number of free parameters by three, we cannot say that there has been an
improvement with respect to the solution presented in Table\,\ref{T2}. 
The inclination of the orbits therefore remain unknown, as do their true masses.

The residuals around the best fit solution are small, but remain slightly
larger than the internal errors (Fig.\,\ref{F1}).
We may then ask if there are other companions with different orbital periods.
For this purpose, we used a genetic algorithm, since we were unable to 
clearly isolate any significant peak in the frequency analysis of the residuals.
The inclusion of additional companions in the system allows us to reduce
the value of $ \sqrt{\chi^2} $ slightly, although this can be justified as a natural consequence of
increasing the number of free parameters. 
Identical adjustments can be obtained with many orbital periods, as
different as 5 or $ 18$~days, frequently with very high eccentricity values.
Therefore, no other companion can be conclusively detected in the
residuals from the orbital solution listed in Table\,\ref{T2}.
The best fit solution was obtained by adding a linear drift to the data, with
$ slope = -0.86 \pm 0.09 \;\mathrm{m\,s}^{-1}$/yr,
allowing us to reduce the value of $ \sqrt{\chi^2} $ to 2.42 and the
$rms$\,=\,$1.21\,\mathrm{m/s}$, 
while the orbital parameters of the two planets remain nearly the same.
The solution in Table\,\ref{T2} must then be considered to be the best
determination achievable so far, and a longer tracking of the system will
provide more accurate orbital parameters for the {HD}\,45364 system.

  

\section{Dynamical analysis}
\label{dynevol}

We now briefly analyze the dynamics and stability of the
planetary system given in Table\,\ref{T2}.
Due to the two planets' proximity and high values of the masses, we expect
that both planets are affected by strong planetary perturbations from each other.
The present orbits of the two planets almost cross (Fig.\,\ref{F6}), and
unless a resonant mechanism is present to avoid close encounters, the system cannot be
stable. 

\subsection{The 3:2 mean motion resonance}

\begin{figure}
    \includegraphics*[width=9.cm]{\figpath 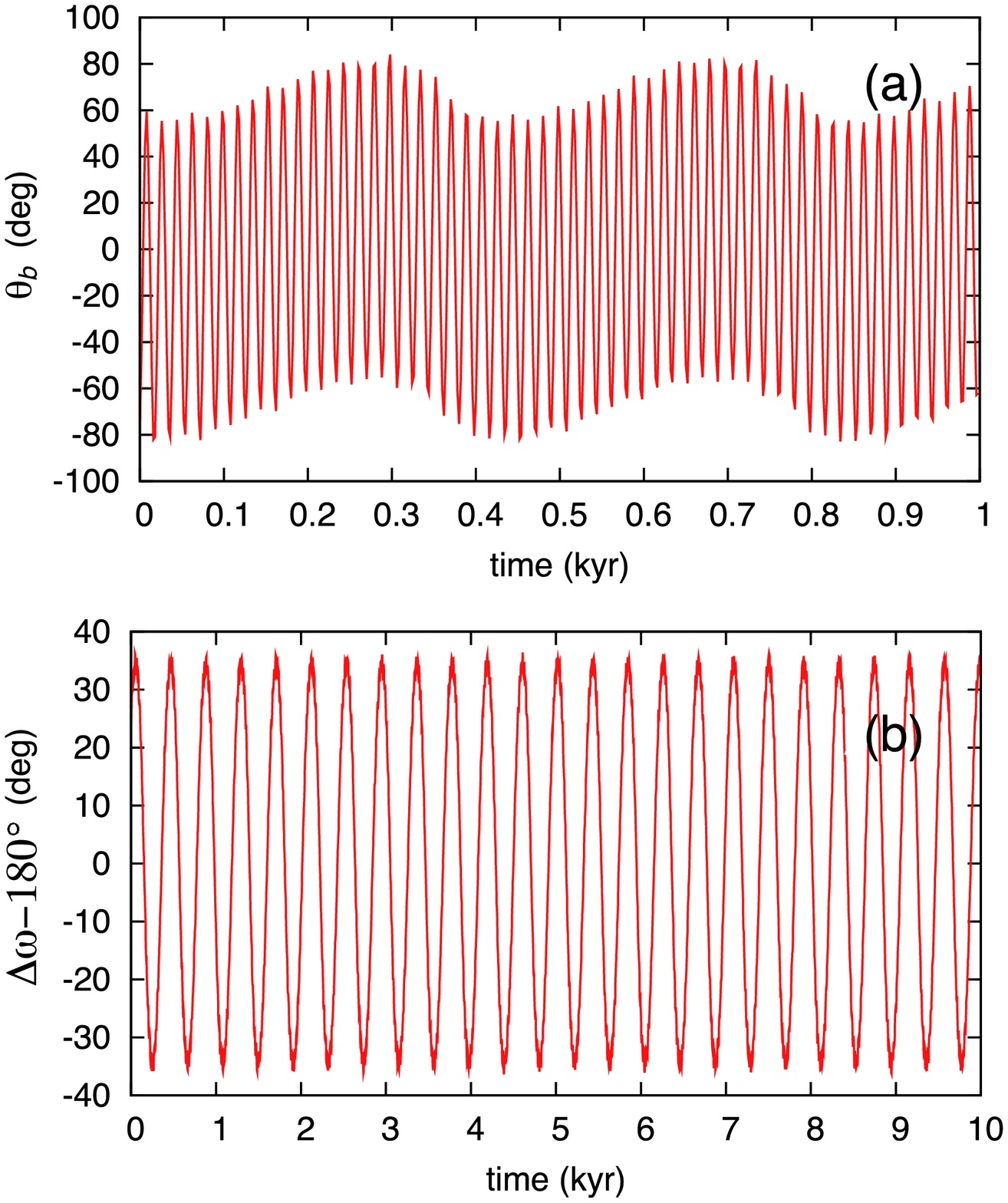} 
  \caption{Variation in the resonant argument, $\theta_b = 2 \lambda_b - 3
  \lambda_c + \omega_b $ (a) and in the secular  argument, $ \Delta \omega =
  \omega_b - \omega_c $ (b), with time. $ \theta_b $ is in libration
  around $ 0^\circ $, with a libration period $P_{l_\theta} \simeq 18.16
  $\,yr, and a principal amplitude of about $ 68.4^\circ $ (Table\,\ref{T4}). 
  $ \Delta \omega $ is in libration around $ 180^\circ $, with a libration
  frequency $ g_{\Delta \omega} = g_1 - g_2 $ (corresponding to a period
  $P_{\Delta \omega} \simeq 413.9 $\,yr), and a maximum amplitude of about $
  36.4^\circ $. \label{F3}}   
\end{figure}

\begin{figure*}
  \centering
    \includegraphics*[width=17cm]{\figpath 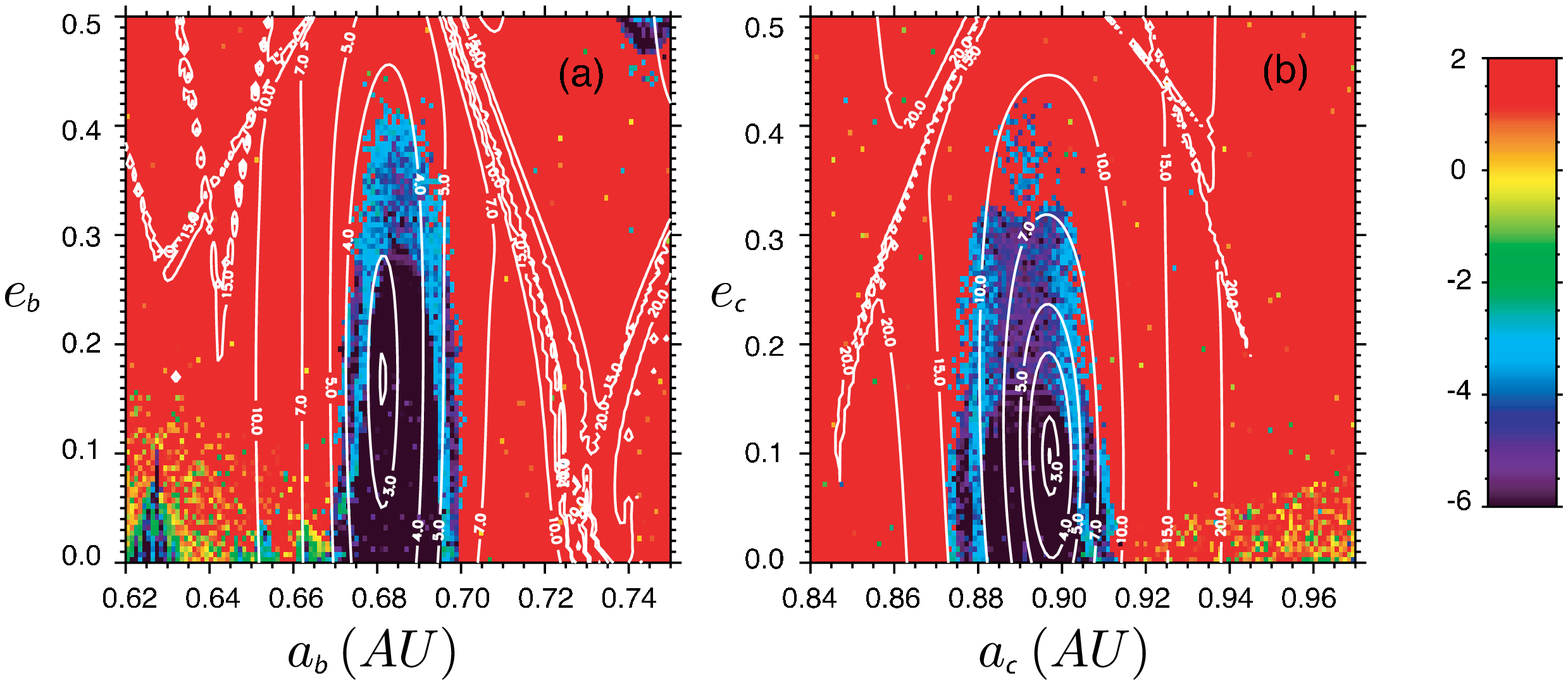} 
  \caption{
  Stability analysis of the nominal fit (Table\,\ref{T2}) 
  of the {HD}\,45364 planetary system. For a fixed initial condition 
  of the outer (a) and inner planet (b), the phase space of the 
  system is explored by varying the semi-major axis $a_k$ and eccentricity 
  $e_k$ of the other planet, respectively. The step size is $0.001$ AU in semi-major axis 
  and $0.005$ in eccentricity. For each initial condition, the 
  full system is integrated numerically over 10~kyr and a stability criterion 
  is derived  with the frequency analysis of the mean longitude
  \citep{Laskar_1990,Laskar_1993PD}.
  As in \citet{Correia_etal_2005}, the chaotic diffusion 
  is measured by the variation in the frequencies. The ``red'' zone corresponds to highly unstable 
  orbits, while  the ``dark blue'' region can be assumed to be stable on a
  billion-years timescale.
  The contour curves indicate the value of $\chi^2$ 
  obtained for each choice of parameters.
  It is remarkable that in the present fit, there is perfect 
  correspondence between the  zone of minimal $\chi^2$ 
  and the 3:2 stable resonant zone, in ``dark blue''.
  \label{F4}}   
\end{figure*}

\begin{table}
 \caption{Fundamental frequencies for the orbital solution in
 Table\,\ref{T2}. 
 \label{T3}} 
 \begin{center}
 \begin{tabular}{crr}
 \hline\hline
      & Frequency   & Period \\
      & $^\circ/yr$ & yr \\
 \hline
 $n_b$ & 576.429624 & 0.624534 \\
 $n_c$ & 384.033313 & 0.937419 \\
 $g_1$ &         $-$0.759309 &  474.115 \\
 $g_2$ &            0.110472 &  3258.74 \\
 $l_\theta$     &  19.820696 &   18.1628 \\ \hline
\end{tabular}
\end{center}
$n_b$ and $n_c$ are the mean motions, $g_1$ and $g_2$ are the
 secular frequencies of the periastrons, and $l_\theta$ is the libration
 frequency of the resonant angle $\theta_b = 2 \lambda_b - 3 \lambda_c +
 \omega_b $. Indeed, we have $ 2 n_b - 3 n_c + g_1 = 0 $.
\end{table}


\begin{table}
\caption{Quasi-periodic decomposition of the resonant angle $\theta_b = 2 \lambda_b - 3
  \lambda_c + \omega_b $ for an integration  over 100~kyr of the orbital solution in
  Table\,\ref{T2}. \label{T4}} 
  \begin{center}
    \begin{tabular}{rrrrrrrr}
\hline\hline
   \multicolumn{5}{c}{\textbf{Combination}} & \multicolumn{1}{c}{$\nu_i$} 
    &\multicolumn{1}{c}{$A_i$}  & \multicolumn{1}{c}{$\phi_i$}  \\
\multicolumn{1}{c}{$n_b$}  & \multicolumn{1}{c}{$n_c$}& 
 \multicolumn{1}{c}{$g_1$}  & \multicolumn{1}{c}{$g_2$} 
    &\multicolumn{1}{c}{ $l_\theta$}     & \multicolumn{1}{c}{(deg/yr)}
    &\multicolumn{1}{c}{(deg)}                & \multicolumn{1}{c}{(deg)}  \\
\hline
   0 &  0 &  0 &  0 &  1 &    19.8207 &        68.444 &     -144.426 \\ 
   0 &  0 & -1 &  1 &  0 &     0.8698 &        13.400 &      136.931 \\ 
   0 &  0 &  1 & -1 &  1 &    18.9509 &         8.606 &      168.643 \\ 
   0 &  0 & -1 &  1 &  1 &    20.6905 &         8.094 &       82.505 \\ 
   0 &  0 & -2 &  2 &  0 &     1.7396 &         2.165 &     -176.138 \\ 
   0 &  0 & -2 &  2 &  1 &    21.5603 &         0.622 &      -50.564 \\ 
   0 &  0 &  0 &  0 &  3 &    59.4621 &         0.540 &      -73.279 \\ 
   1 & -1 &  0 &  0 & -1 &   172.5756 &         0.506 &        7.504 \\ 
   1 & -1 &  0 &  0 &  0 &   192.3963 &         0.501 &      -46.923 \\ 
   0 &  0 & -3 &  3 &  0 &     2.6093 &         0.416 &     -129.207 \\ 
   0 &  0 &  2 & -2 &  1 &    18.0811 &         0.420 &      121.712 \\ 
   1 & -1 &  0 &  0 &  1 &   212.2170 &         0.416 &       78.651 \\ 
   0 &  1 & -1 &  0 &  0 &   384.7926 &         0.451 &      176.155 \\ 
   1 & -1 &  0 &  0 & -2 &   152.7549 &         0.424 &     -118.070 \\ 
   0 &  1 & -1 &  0 & -1 &   364.9719 &         0.341 &       50.581 \\ 
   0 &  1 & -1 &  0 &  1 &   404.6133 &         0.274 &      121.729 \\ 
   0 &  0 &  1 & -1 &  3 &    58.5923 &         0.212 &     -120.210 \\ 
   1 & -1 &  0 &  0 &  2 &   232.0377 &         0.201 &       24.225 \\ 
   0 &  0 & -1 &  1 &  3 &    60.3319 &         0.211 &      153.652 \\ 
   1 &  0 & -1 &  0 & -1 &   557.3682 &         0.182 &      -86.342 \\ 
  \hline
\end{tabular}
\end{center}
  We have $\theta_b = \sum_{i=1}^N A_i \cos(\nu_i\, t + \phi_i)$, where 
  the amplitude and phases $A_i$, $\phi_i$ are given in degree, and the 
  frequencies $\nu_i$ in degree/year. 
  We only give the first 20 terms, ordered by
  decreasing amplitude. All terms are identified as  integer combinations 
  of the fundamental frequencies given in
  Table\,\ref{T3}. The fact that we are able to express all the main frequencies of $
  \theta_b $ in terms of exact combinations of the fundamental frequencies $ g_1 $, $
  g_2 $ and $ l_\theta $ is a signature of a very regular motion. 
\end{table}

The ratio between the orbital periods of the two planets determined 
by the fitting process (Table\,\ref{T2}) is $ P_c
/ P_b = 1.511 $, suggesting that the system may be trapped in a 3:2
mean motion resonance.
To test the accurancy of this scenario, we performed a frequency analysis of the 
orbital solution listed in Table\,\ref{T2} computed over 100~kyr.
The orbits of the planets are integrated with the 
symplectic integrator SABAC4 of \citet{Laskar_Robutel_2001}, using a step size
of 0.02~years.
We conclude that in the nominal solution of Table\,\ref{T2},
the two planets in the {HD}\,45364 system indeed show a
3:2 mean motion resonance, with resonant argument:
\begin{equation}
\theta_b = 2 \lambda_b - 3 \lambda_c + \omega_b \ . \label{RArg}
\end{equation}
The fundamental frequencies of the systems are 
the two mean motions $n_b$ and $n_c$, the two secular frequencies of the 
periastrons $g_1$ and $g_2$, and the libration frequency of the resonant argument
$l_{\theta}$ (Table\,\ref{T3}).
These frequencies are not independent because, due to the 3:2 resonance, 
we have up to the precision of the determination of the frequencies ($\approx 10^{-10}$), 
\begin{equation}
2 n_b-3 n_c + g_1 = 0 \ .
\end{equation}

\begin{figure*}
  \centering
    \includegraphics*[width=15.cm]{\figpath 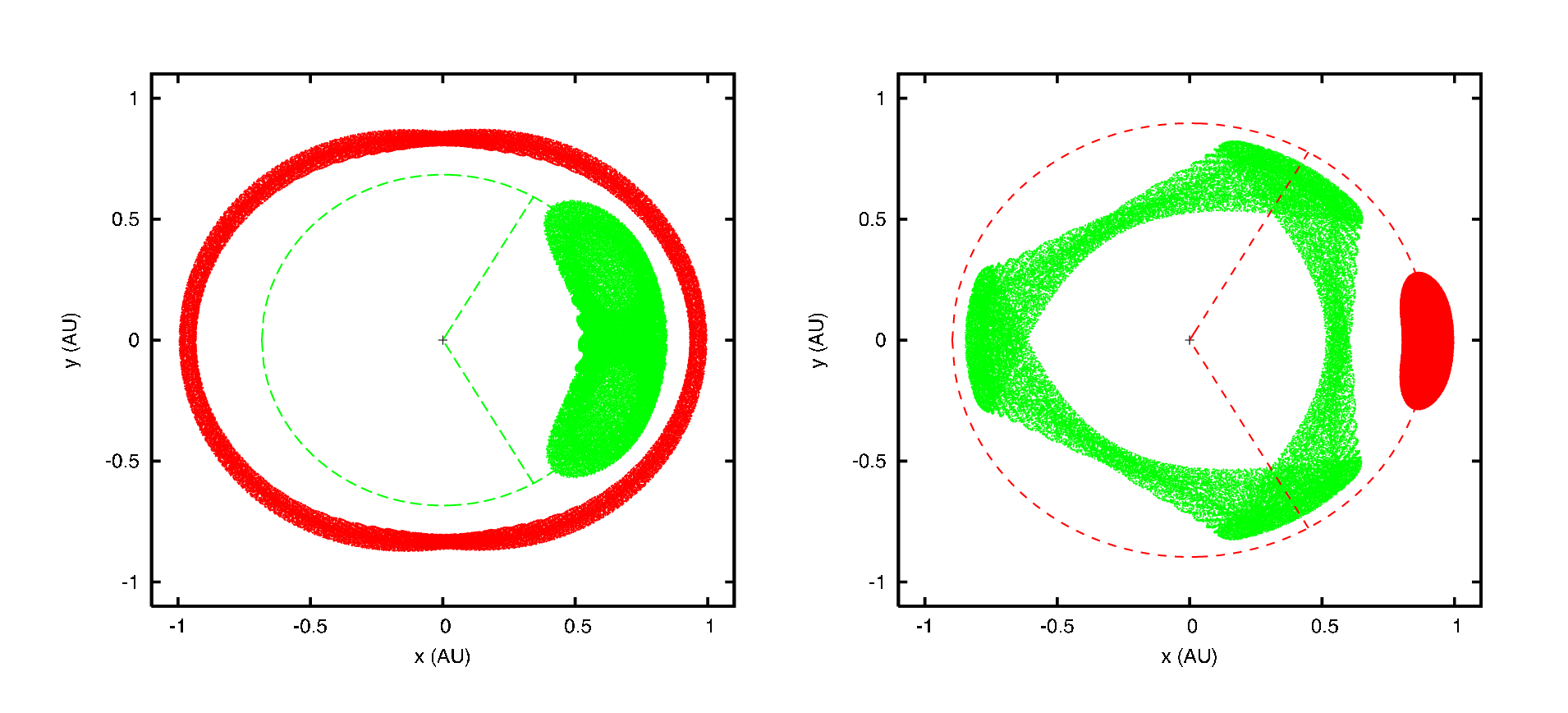} 
  \caption{Evolution of the {HD}\,45364 planetary system over 100~kyr in the
  co-rotating frame of the inner planet (left) and outer planet (right). $x$ and
  $y$ are spatial coordinates in a frame centered on the star and rotating
  with the frequency $n_b$ (left) or $n_c$ (right). 
  Due to the 3:2 mean motion resonance the planets never get too close, the minimal distance of approach
  being 0.371~AU, while the maximum distance can reach 1.811~AU. We also observe
  that the trajectories are repeated every 3 orbits of the inner planet and
  every 2 orbits of the outer planet. \label{F5}}   
\end{figure*}

The resonant argument $ \theta_b $ is in libration around $0^\circ $,
with a libration period $ 2 \pi / l_\theta = 18.16 $~yr, and an associated 
amplitude of about 68.44 degrees (Fig.\,\ref{F3}a, Table\,\ref{T4}). 
For the complete solution, the libration amplitude can reach
more than 80 degrees because additional periodic terms are present. 
In Table\,\ref{T4}, we provide a quasi-periodic  decomposition of the resonant angle $ \theta_b $
in terms of decreasing amplitude.
All the quasi-periodic terms are easily identified as integer combinations of the 
fundamental frequencies (Table\,\ref{T3}).
Since the resonant angle is modulated by a relatively short period of about 18~years, the
observation of the system over a few additional decades may provide an
estimate of the libration amplitude and thus a strong constraint
on the parameters of the system.

Although the mean motions $ n_b $ and $ n_c $ can be associated with the two
planets $b$ and $c$, respectively, it is not the case for the secular 
frequencies $g_1$ and $g_2$, and incidentally, both periastrons precess with mean 
frequency $g_1$ that is retrograde, with a period of 474.115 years. The two
periastrons are thus 
locked in an antipodal state, and the difference $\Delta \omega = \omega_b-\omega_c$ is in libration 
around $180^\circ$ with an amplitude of about $36.4^\circ$ (Fig.\,\ref{F3}b).
As a result, the argument  $\theta_c = 2 \lambda_b - 3 \lambda_c + \omega_c$ 
librates around $180^\circ$ with the same libration frequency
$l_{\theta}$. 

\subsection{Stability analysis}

\begin{figure}
    \includegraphics*[width=8.5cm]{\figpath 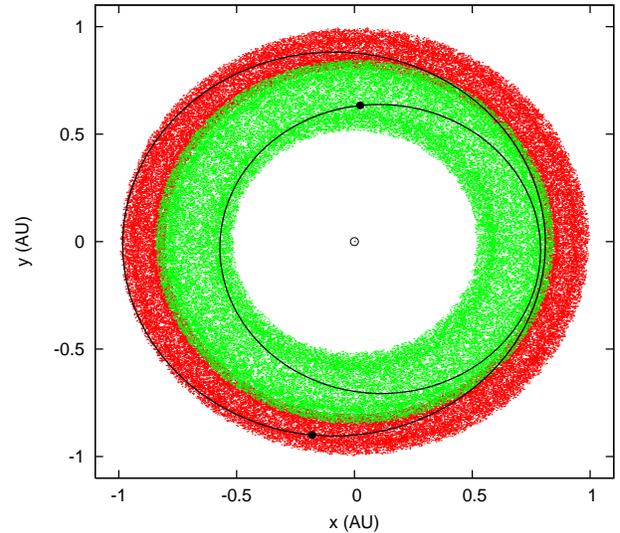} 
  \caption{Long-term evolution of the {HD}\,45364 planetary system over 5~Gyr
  starting with the orbital solution from Table\,\ref{T2}.  
  The panel shows a face-on view of the system. $x$ and $y$ are spatial
  coordinates in a frame centered on the star. Present orbital solutions are traced
  with solid lines and each dot corresponds to the position of the planet every
  100~kyr. The semi-major axes (in AU) are almost constant ($ 0.676 < a_b < 0.693 $
  and $ 0.893 < a_c < 0.902 $), but the eccentricities undergo significant variations
  ($ 0.123 < e_b < 0.286 $ and $ 0.042 < e_c < 0.133 $). The fundamental periods
  related to the precession of the periastrons are $ 2 \pi / g_1 = - 474$~yr and 
  $ 2 \pi / g_2 = 3259 $~yr. 
  \label{F6}}   
\end{figure}

To analyze the stability of the nominal solution and
confirm the presence of the 3:2 resonance, 
we performed a global frequency analysis \citep{Laskar_1993PD} 
in the vicinity of the nominal solution (Fig.\,\ref{F4}), in the same way as
achieved for the {HD}\,202206 system by \citet{Correia_etal_2005}.
For each planet, the system is integrated on a regular 2D mesh of initial conditions, 
with varying semi-major axis and eccentricity, while the other parameters are 
retained at their nominal values. The solution is integrated over 10~kyr 
for each initial condition and a stability indicator is computed 
to be the variation in the measured mean motion over the two consecutive 
5~kyr intervals of time. For regular motion, there is no significant variation 
in the mean motion along the trajectory, while it can vary significantly 
for chaotic trajectories. The result is reported in color in Fig.\,\ref{F4}, 
where ``red'' represent the strongly chaotic trajectories, and ``dark blue'' 
the extremely stable ones. In  both plots (Figs. \ref{F4}a,b), 
it appears that the only stable zone that exists in the vicinity of the 
nominal solution are the stable 3:2 resonant zones. 

It is quite remarkable that, in contrast to the findings of
\citet{Correia_etal_2005} for the 5:1 resonance in 
the {HD}\,202206 system, there is perfect coincidence between
the stable 3:2 resonant islands, and curves of minimal $\chi^2$
obtained in comparing with the observations.
Since these islands are the only stable zones in the vicinity, this
picture presents a very coherent view of dynamical analysis and radial
velocity measurments, which reinforces the confidence that the present system 
is in a 3:2 resonant state.  

In Fig.\,\ref{F5}, we plot the evolution of the {HD}\,45364 planetary system over
100~kyr in the rotating frame of the inner and the outer planet, respectively.
Due to the 3:2 mean motion resonance trapping, the relative positions of the two
planets are repeated 
and never become closer than about
0.37~AU, preventing close encounters and the consequent destruction of the system.
The paths of the planets in the rotating frame illustrate the relationship between the
resonance and the frequency of conjunctions with the internal or external planet.
The inner planet is in a 3:2 resonance with the outer planet, so the orbital
configuration of the system is repeated every 3 orbits of the inner planet and
every 2 orbits of the outer planet.
In this particular frame, we are also able to see the libration of each planet
around its equilibrium position.

\subsection{Orbital evolution}

From the previous stability analysis, it is clear that the 
 {HD}\,45364 planetary system listed in Table\,\ref{T2} is trapped in 
 a 3:2 mean motion resonance  and stable over a Gyr timescale.
Nevertheless, we tested directly this by performing a numerical integration
of the orbits over 5~Gyr using the symplectic integrator SABAC4 of
\citet{Laskar_Robutel_2001} with a step size of 0.02~years.
The results displayed in Fig.\,\ref{F6} show that the orbits indeed 
evolve in a regular way, and remain stable throughout the simulation,
which corresponds to the estimated age of the star. 

Because of the strong gravitational interactions between the two planets, both orbital
eccentricities present significant variations.
The eccentricity of the inner planet is within $ 0.12 < e_b < 0.29 $, while that
of the outer planet is within $ 0.04 < e_c < 0.13 $.
We also observe rapid secular variations in the orbital parameters, mostly driven by
the rapid secular frequency $g_1$, of period $ 2 \pi / g_1 \approx 474 $~yr
(Table\,\ref{T3}).
These secular variations in the orbital elements occur much more rapidly than in
our Solar System, which should enable them to be detected directly from
observations.


\section{Discussion and conclusion}

We have reported the detection of two planets orbiting the star
{HD}\,45364, with orbital periods of 228 and 342~days, and minimum
masses of 0.187 and 0.658\,$M_\mathrm{Jup}$, respectively.
A dynamical analysis of the system has further shown a 3:2 mean motion resonance between
the two planets, which ensures its stability over 5\,Gyr despite the proximity of
the two orbits.
This is the first time that such an orbital resonant configuration has been observed for
extra-solar planets, although an analogue does exist in our Solar System
composed by Neptune and Pluto.
However, while Neptune evolves in an almost circular orbit and is much more massive
than Pluto (which is the largest member of the asteroid family of the Plutinos), the
two planets around {HD}\,45364 have masses comparable to those of Saturn and
Jupiter, and are evolving in orbits with moderate eccentricity.

Dynamically, the system is extremely interesting. 
In the nominal solution, the resonant angle $ \theta_b =  2 \lambda_b - 3 \lambda_c + \omega_b $ is in libration
around $ 0^\circ $, with a libration period of 18.16~years
and a dominating amplitude of 68.44 degrees.
Such an orbital configuration may have been reached through the dissipative process
of planet migration during the early stages of the system evolution.  
However, after a resonant capture, subsequent migration produces a significant
increase in planetary eccentricities, unless a damping mechanism is applied.
Since the eccentricities of the two planets around {HD}\,45364 are relatively
small (Table\,\ref{T2}), migration may cease shortly after capture in resonance
occurs, or, according to \citet{Crida_etal_2008}, an inner disc must be present.
This singular planetary system may then provide important constraints on planetary
formation and migration scenarios.

The strong gravitational interactions between the planets may also allow us to
model their effect more accurately in the near future.
With the current precision of {\small HARPS}, about 1~m/s for {HD}\,45364, we
expect to detect the signature of planet-planet interactions in data in a few
decades.
The planet-planet interactions may provide important information about the
inclination of the orbital planes and allow us to determine the precise masses
of both planets.


\begin{acknowledgements}
We acknowledge support from the Swiss National Research Found (FNRS), the Geneva
University, Funda\c{c}\~{a}o Calouste Gulbenkian (Portugal), and French CNRS.
\end{acknowledgements}

\bibliographystyle{aa}
\bibliography{correia}

\end{document}